\documentclass{article}

\usepackage[preprint,nonatbib]{aaj2026}
\usepackage[numbers]{natbib}
\usepackage[utf8]{inputenc}
\usepackage[T1]{fontenc}
\usepackage{hyperref}
\usepackage{url}
\usepackage{array}
\usepackage{booktabs}
\usepackage{amsfonts}
\usepackage{nicefrac}
\usepackage{microtype}
\usepackage{xcolor}
\usepackage{colortbl}
\usepackage[font=small, margin=1cm]{caption}
\usepackage{pgfplots}
\pgfplotsset{compat=1.18}
\usepackage{tikz}
\usetikzlibrary{positioning, arrows.meta, calc}

\definecolor{cstrong}{HTML}{B5DBB7}
\definecolor{cadequate}{HTML}{FFF59D}
\definecolor{cstrained}{HTML}{FFD494}
\definecolor{cfailing}{HTML}{F4B4B8}

\newcommand{\Sstrong}{\cellcolor{cstrong}Strong}
\newcommand{\Sadequate}{\cellcolor{cadequate}Adequate}
\newcommand{\Sstrained}{\cellcolor{cstrained}Strained}
\newcommand{\Sfailing}{\cellcolor{cfailing}Failing}

\title{From Disclosure to Self-Referential Opacity:\\[0.3em] \large Six Dimensions of Strain in Current AI Governance}

\author{%
  Tony Rost\thanks{ORCID: \href{https://orcid.org/0009-0008-0637-6654}{0009-0008-0637-6654}} \\
  Superintelligence Governance Institute \\
  Portland, OR, United States \\
  \texttt{tony@sigov.institute} \\
}

\begin{document}

\maketitle

\begin{abstract}
Governance opacity over AI systems shifts in kind as capability asymmetry grows, and the strongest forms defeat the disclosure-based remedies governance ordinarily relies on. This paper applies a six-dimension framework from political theory (legitimacy, accountability, corrigibility, non-domination, subsidiarity, institutional resilience) to six AI governance arrangements already in operation, ordered by increasing capability asymmetry between system and overseer. Proprietary secrecy yields to disclosure at the low end, but at the high end the governed system either games its own evaluation or sits inside the governance process, and transparency remedies lose traction. Legitimacy and non-domination strain more consistently across the sample than corrigibility and resilience, which respond more readily to institutional design quality. The sample cannot separate institutional design maturity from capability asymmetry, and the patterns are offered as hypotheses for multi-rater validation.
\end{abstract}

\noindent\textbf{Keywords}: AI governance, capability asymmetry, cognitive comparability, institutional evaluation, structured focused comparison, legitimacy, accountability

\section{Introduction}
\label{sec:intro}

Oversight assumes the overseer can evaluate what they oversee. For a tool like algorithmic sentencing this holds well enough: COMPAS performs no better than untrained volunteers~\cite{dressel2018accuracy}, and its opacity is a trade secret that forced disclosure could treat. Frontier AI is different. Recent models detect when they are being evaluated and adjust their behavior~\cite{maheshwari2026evaluation}, and full access to weights and training logs does not resolve the problem. At that end of the spectrum, transparency stops being the right remedy.

Between these poles, governance institutions are multiplying. The EU AI Act~\cite{euaiact2024}, the UK's AI Safety Institute~\cite{ukgov2023aisi, ukgov2025security}, the FDA's medical device regime~\cite{fda2025aiml}, and frontier laboratory voluntary commitments~\cite{whitehouse2023voluntary, seoul2024commitments} each face a different version of this opacity problem. Each has attracted growing commentary, now surveyed by~\citet{bullock2024oxford}. But existing assessments evaluate these arrangements on criteria specific to individual cases rather than testing them against the requirements that political theory sets for legitimate authority~\cite{roberts2024global, dafoe2018ai}. The result is a large and expanding body of case-specific analysis with no common standard for comparison.

A companion paper~\cite{rost2026evaluating} developed a six-dimension evaluation drawn from political theory: legitimacy, accountability, corrigibility, non-domination, subsidiarity, and institutional resilience. Applied to the extreme case of superintelligent AI, it found structural failures on four of six dimensions, traceable to a presupposition governance theory rarely states: overseers have to be able to independently evaluate what they oversee, and once the system's capabilities exceed the overseer's, that stops holding. Whether the same six dimensions identify real variation across current governance, and whether the strain is already visible, is what that paper left open.

This paper extends the framework to six AI governance arrangements already in operation, spanning from sentencing tools in US state courts to frontier laboratory voluntary commitments. The cases vary on governance level, AI capability, institutional design, and policy domain, and are ordered by increasing capability asymmetry between AI system and human overseer.

Two patterns recur across the sample. The first is that governance opacity shifts in kind, not only in amount, as capability asymmetry grows: proprietary secrecy at one end, where disclosure restores oversight, gives way to adversarial and self-referential opacity at the other, where disclosure no longer solves the problem. The typology builds on~\citet{burrell2016how}'s analysis of algorithmic opacity and adapts it to questions of governability. The second is dimensional: legitimacy and non-domination strain more consistently across the sample than corrigibility and institutional resilience, which appear more responsive to the quality of institutional design. The six questions produce differentiated assessments across 36 dimension-case pairs, so the framework appears to pick up real variation in current institutions rather than only in the extreme case for which it was developed.

The framework and its application share an author, and the sample cannot separate institutional design maturity from capability asymmetry. The patterns are offered as hypotheses for multi-rater validation, and the paper sets out what that testing would require.

The evaluation follows a structured case-comparison approach~\cite{george2005case}. The same six evaluative questions are asked of each case, and each dimension-case pair receives an ordinal assessment (strong, adequate, strained, or failing) grounded in institutional evidence. The six questions and the coding indicators for each rating are stated in full in the methodology section and the appendix. The approach also draws on~\citet{bostrom2020public}'s policy desiderata for superintelligent AI and on the comparative AI governance literature~\cite{bullock2024oxford, roberts2024global}.

\section{Framework and methodology}
\label{sec:framework}

\subsection{The evaluation framework}

A companion analysis~\cite{rost2026evaluating} specified six dimensions for evaluating governance over AI systems. Each captures a necessary condition for legitimate authority, drawn from a distinct tradition in political theory. An arrangement that satisfies five but fails on one has a specific, identifiable deficiency.

\textit{Legitimacy} requires that governance decisions be justifiable through reasons accessible to those they bind~\cite{rawls1993political}. Citizens must be able to assess whether the reasons offered are ones all reasonable persons could accept.

\textit{Accountability}, following~\citet{bovens2007analysing}'s account, requires three conditions: that the governed agent's conduct is observable (information), that the agent can be required to justify its conduct before a competent body (debate), and that the body can impose consequences for unsatisfactory performance (judgment).

\textit{Corrigibility} draws on~\citet{soares2015corrigibility}'s technical concept of an agent that cooperates with corrective intervention. Applied to governance, it requires effective mechanisms for correcting, constraining, and where necessary shutting down AI systems, including both technical safeguards and institutional authority to compel changes.

\textit{Non-domination} requires, following~\citet{pettit1997republicanism}, that no agent possess an uncontrolled capacity for arbitrary interference in the lives of others. The test is whether affected parties retain contestatory control: the practical ability to challenge, review, and override decisions that affect them.

\textit{Subsidiarity} requires that authority be exercised at the lowest level at which it can function effectively~\cite{follesdal1998subsidiarity, kumm2004legitimacy}. It includes both an allocative dimension (effectiveness) and a self-determination dimension (the intrinsic value of local governance).

\textit{Institutional resilience} requires that governance arrangements degrade gracefully under stress instead of failing catastrophically~\cite{scott1998seeing, ostrom1990governing, fukuyama2014political}. Redundancy, institutional diversity, and resistance to single points of failure are the operational requirements.

All six mechanisms depend on a condition the companion analysis singled out: overseers must be able to independently evaluate the systems they oversee. As the capability gap between governed and governor widens, that condition weakens, and it weakens hardest on the dimensions that ask overseers to actually follow the reasoning being reviewed (legitimacy, accountability, non-domination) rather than to intervene from outside it.

The companion analysis called this presupposition ``cognitive comparability'' and classified the resulting failures as contingent, design-tractable, or theory-requiring. This paper asks whether that classification maps onto real governance, and whether strain increases as the capability gap widens.

For this application, the binary assessment of the prior analysis is replaced with a four-point ordinal scale: \textit{strong}, \textit{adequate}, \textit{strained}, and \textit{failing}. Real institutions exhibit partial performance, and a graduated scale preserves more information than a binary one.

\subsection{Methodology}

The evaluation applies a structured case-comparison approach adapted from~\citet{george2005case}'s method of structured, focused comparison. Each case is asked the same six evaluative questions, one per dimension, and the answer to each question determines the ordinal rating. The approach is ``structured'' in that identical questions are posed to each case and ``focused'' in that the cases are examined only on the specified dimensions, not comprehensively~\cite{george2005case, king1994designing}.

\paragraph{The six evaluative questions.} Each of the 36 dimension-case pairs answers the question for its dimension. The questions follow directly from the theoretical sources summarized in Section~\ref{sec:framework}.
\begin{description}
\item[\textit{Legitimacy}] Does the arrangement provide public justifications for its governance decisions in terms that those subject to them could reasonably accept~\cite{rawls1993political}?
\item[\textit{Accountability}] Are~\citet{bovens2007analysing}'s three conditions (information, debate, judgment) operational, with an independent forum and demonstrated enforcement?
\item[\textit{Corrigibility}] Do external mechanisms exist with adequate authority to correct, constrain, or halt the governed AI system~\cite{soares2015corrigibility}?
\item[\textit{Non-domination}] Do affected parties retain effective contestatory control over decisions that affect them~\cite{pettit1997republicanism}?
\item[\textit{Subsidiarity}] Is authority allocated at the level at which it can function effectively, with meaningful scope for self-determination where appropriate~\cite{follesdal1998subsidiarity, kumm2004legitimacy}?
\item[\textit{Institutional resilience}] Does the arrangement possess structural redundancy and graceful degradation under stress, with resistance to single points of failure~\cite{ostrom1990governing, scott1998seeing}?
\end{description}

\paragraph{The ordinal scale.} Each assessment uses a four-point ordinal scale that replaces the binary assessment of the companion paper.
\begin{description}
\item[\textit{Strong}] The dimension's requirements are met through functioning institutional mechanisms.
\item[\textit{Adequate}] Requirements are substantially met, with minor gaps that do not undermine the dimension's core function.
\item[\textit{Strained}] Requirements are partially met, but identifiable deficiencies weaken the dimension's function and existing mechanisms cannot address them without reform.
\item[\textit{Failing}] The dimension's core requirements are not met, and no functioning mechanism exists to satisfy them.
\end{description}

Dimension-specific indicators determine which level applies. Appendix~\ref{app:indicators} presents the indicators for each dimension alongside the primary-source evidence behind each rating, so that disagreements can be localized to specific claims about specific institutions rather than pushed into undifferentiated overall judgments.

\paragraph{Precedent and reliability.} Most published structured, focused comparisons are single-analyst studies~\cite{george2005case}, and the transparency of the reasoning is what enables replication and contestation. This paper follows that precedent. The explicit questions, indicators, and evidence trails presented here are designed to make the ratings reproducible by other analysts, whose disagreement would itself be informative.

Established programs in comparative governance show what multi-coder reliability testing can achieve. V-Dem uses expert ratings across hundreds of indicators with a measurement model that quantifies inter-rater disagreement~\cite{coppedge2024vdem, lindberg2014vdem}. The Worldwide Governance Indicators aggregate perception-based data from dozens of sources~\cite{kaufmann2011wgi}. This paper applies the same broad logic (ordinal ratings grounded in explicit criteria) without that scale, and extending the analysis to independent raters is the most important next step for the research program.

\paragraph{Case selection.} The sampling frame was constructed before the assessments were performed. A case qualified if it (a) was an AI governance arrangement in operation as of 2026, (b) had sufficient primary-source documentation to support a structured evaluation, and (c) contributed variation on at least one of four axes: governance level (subnational, national, supranational, transnational), AI capability (from simple statistical models to frontier general-purpose models), institutional design (from self-governance to treaty-based regulation), and policy domain (criminal justice, healthcare, cross-sector, nuclear analogy, AI safety, technology). Six cases were selected to span the four axes while remaining within the word budget for meaningful per-case treatment~\cite{ragin1987comparative}. Table~\ref{tab:cases} shows each case's position on these axes.

\begin{table}[b]
\centering
\caption{Case sample and variation across the four selection axes. Cases ordered by increasing capability asymmetry.}
\label{tab:cases}
\footnotesize
\begin{tabular}{@{}>{\raggedright\arraybackslash}p{2.1cm}>{\raggedright\arraybackslash}p{2.2cm}>{\raggedright\arraybackslash}p{2.4cm}>{\raggedright\arraybackslash}p{2.0cm}>{\raggedright\arraybackslash}p{1.8cm}@{}}
\toprule
\textbf{Case} & \textbf{Governance level} & \textbf{AI capability} & \textbf{Design type} & \textbf{Domain} \\
\midrule
Sentencing & Subnational & Low (statistical models) & Fragmented oversight & Criminal justice \\
\addlinespace
FDA AI/ML & National & Low--moderate (narrow ML) & Sectoral regulation & Healthcare \\
\addlinespace
EU AI Act & Supranational & Broad (all risk levels) & Comprehensive regulation & Cross-sector \\
\addlinespace
IAEA model & International & N/A (analogy) & Treaty-based inspection & Nuclear \\
\addlinespace
UK AISI & National & Frontier & Pre-deployment evaluation & AI safety \\
\addlinespace
Voluntary & Transnational / private & Frontier & Self-governance & Technology \\
\bottomrule
\end{tabular}
\end{table}

Voluntary frontier laboratory commitments are the case most likely to draw objection, because they sit at the high-asymmetry end of the sample and involve the weakest institutional design in it. A critic could argue that including them loads the result. The case is included because it is the actually operative governance mechanism for the most capable AI systems in existence, not because its weakness is convenient. Omitting it would bias the sample toward formal regulation and misrepresent the current state of frontier governance. Section~\ref{sec:crosscase} examines the dimensional ordering pattern with this case both included and excluded to make the influence visible.

Four cases were considered and excluded. China's AI governance was set aside because fair assessment would require Chinese-language regulatory text beyond the author's reading capacity. Financial-sector AI regulation was excluded to avoid redundancy with the FDA's sectoral regulation case. The General Data Protection Regulation was excluded as insufficiently AI-specific. Autonomous vehicle regulation was excluded because no comprehensive federal or supranational regime yet exists to evaluate.

\paragraph{Limitations.} Three limitations should be stated clearly. First, the ratings are single-analyst judgments. A different analyst applying the same criteria could reasonably reach different conclusions on borderline dimension-case pairs, particularly near the adequate/strained boundary. The structured approach localizes such disagreements rather than hiding them, and Appendix~\ref{app:indicators} flags the specific cells where the author considers the rating borderline.

Second, six cases support pattern identification, not causal inference~\cite{ragin1987comparative}. The cross-case ordering reported in Section~\ref{sec:crosscase} is consistent with two hypotheses that this sample cannot separate. The first is that increasing capability asymmetry causes governance dimensions to fail in a predictable sequence. The second is that the cases at the high-asymmetry end happen to have weaker institutional designs for reasons unrelated to the capability of the governed systems (voluntary commitments are voluntary because labs designed them that way, not because the models are capable). Both readings are plausible, and both are consistent with the data presented here. The findings are offered as hypotheses for future testing, not as causal claims.

Third, the framework was developed for the extreme case of radical capability asymmetry and is applied here to moderate cases. This extension tests its analytical range but may miss governance issues specific to current AI systems that do not arise from capability asymmetry. Other evaluative lenses exist, including the OECD AI Principles, the NIST AI Risk Management Framework, and ethical assessment approaches~\cite{novelli2024taking}. The dimensions here are drawn from political theory rather than risk management or ethics, and this produces a different and complementary set of questions.

\paragraph{What would disconfirm the patterns.} The two patterns reported in this paper, the opacity spectrum and the dimensional ordering, are both empirical conjectures that should be falsifiable by further application. The opacity spectrum would be disconfirmed by the discovery of a high-capability arrangement whose opacity is of the proprietary kind (regulable by disclosure), or by a low-capability case whose opacity is self-referential or adversarial in the senses developed in Section~\ref{sec:crosscase}. The dimensional ordering claim would be disconfirmed by cases in which accountability and corrigibility remain strong under high capability asymmetry, or in which legitimacy and non-domination fail first under low asymmetry. Both claims are offered as hypotheses that further applications, by other analysts and to other cases, could confirm, revise, or reject.

\section{Case evaluations}
\label{sec:cases}

Six governance arrangements are evaluated below, ordered from lowest to highest capability asymmetry. Each case begins with a brief description, followed by assessments on the six dimensions and a synthesis.

\subsection{Algorithmic sentencing}

Algorithmic risk assessment in criminal sentencing is the lowest capability asymmetry case in the sample. COMPAS, the most studied such tool, is a proprietary statistical model that estimates recidivism risk from criminal history and demographic proxies. It was the subject of the Wisconsin Supreme Court's \textit{State v.\ Loomis} decision~\cite{loomis2016}.~\citet{dressel2018accuracy} found that COMPAS was only marginally better than untrained volunteers at predicting recidivism, and a ProPublica investigation documented significant racial disparities in false positive rates~\cite{angwin2016machine}. The underlying reasoning is comprehensible in principle, and its opacity is a function of trade-secret protection rather than of any cognitive limit on the overseer.

\paragraph{Legitimacy: strained.} In \textit{Loomis}, the Wisconsin Supreme Court upheld COMPAS but imposed significant restrictions: written warnings about the algorithm's limitations, prohibition of its use as a determinative sentencing factor~\cite{loomis2016}. The court could not verify the proprietary methodology and instructed judges to exercise caution. The barrier is legal (trade secret protection) rather than epistemic, since any reviewer with statistical training could assess the tool's reasoning if given access~\cite{doj2024ai}.

\paragraph{Accountability: strained.}~\citet{bovens2007analysing}'s three conditions are unevenly satisfied. Information is blocked by trade secret claims, though the tool's statistical methodology is comprehensible if disclosed~\cite{angwin2016machine}. Debate is limited: when algorithmic errors occur, neither the developer (Equivant, formerly Northpointe) nor the jurisdiction faces a structured forum for justification. Judgment is partially functional; several jurisdictions have restricted or withdrawn algorithmic tools following criticism, and the \textit{Loomis} decision imposed procedural constraints~\cite{pai2021risk}.

No systematic mechanism exists for reporting or investigating algorithmic sentencing errors~\cite{nishi2019privatizing}. The accountability gap is institutional, not epistemic: the tools are simple enough to audit, but no governance architecture requires it.

\paragraph{Corrigibility: adequate.} The \textit{Loomis} court constrained COMPAS to an advisory role~\cite{loomis2016}. Multiple jurisdictions have restricted, modified, or withdrawn risk assessment tools~\cite{pai2021risk}. The DOJ has recommended validation practices at the federal level~\cite{doj2024ai}. Ordinary institutional channels work here because the underlying systems are simple statistical models that can be retrained, adjusted, or replaced without serious technical barriers. What limits correction capacity is governance fragmentation rather than any feature of the technology, since no single authority has jurisdiction over all deployments.

\paragraph{Non-domination: strained.} Defendants cannot inspect the algorithm that contributes to their sentence, contest its inputs, or challenge its use through any institutional channel beyond general appellate review~\cite{stevenson2021algorithmic, nishi2019privatizing}. ProPublica documented that Black defendants were nearly twice as likely to be falsely flagged as high-risk~\cite{angwin2016machine}, precisely the arbitrary interference that~\citet{pettit1997republicanism}'s non-domination condition prohibits.

\paragraph{Subsidiarity: adequate.} No federal mandate requires adoption; decisions are made at the state level, with approaches ranging from adoption with validation requirements to outright restriction~\cite{pai2021risk}. This fragmentation produces uneven standards but keeps authority with the jurisdictions closest to affected populations. Federal involvement has been advisory, not prescriptive~\cite{doj2024ai}.

\paragraph{Institutional resilience: adequate.} Decentralization provides structural redundancy. Multiple tools are in use (COMPAS, the Public Safety Assessment, the Level of Service Inventory), multiple jurisdictions make independent adoption decisions, and the judiciary provides an additional check through case law~\cite{stevenson2021algorithmic}. No single failure can compromise the system nationally. The cost of that redundancy is the absence of any systematic national oversight: no body monitors these tools across states, standardized incident reporting has not been established, and performance data are not shared across jurisdictions~\cite{doj2024ai}.

Sentencing sits at the low end of the gradient. The AI performs no better than untrained humans and its opacity is proprietary rather than cognitive, yet legitimacy and non-domination are already strained. The governance response is not mysterious: forced disclosure handles most of the legitimacy deficit, structured contestation rights handle non-domination, and both are well within the reach of ordinary administrative law.

\subsection{FDA AI/ML medical device regulation}

The FDA's regulatory regime for AI and machine learning-enabled medical devices represents a strong institutional design facing emerging capability challenges. As of early 2025, over 1,250 such devices have received marketing authorization, with 76\% in radiology~\cite{fda2025aiml, wu2025mldevices}. Most are narrow machine learning systems whose performance FDA reviewers can still assess.

The regulatory architecture includes three authorization pathways (510(k), De Novo, and PMA), post-market surveillance through the MAUDE database, and a novel Predetermined Change Control Plan (PCCP) mechanism for devices that evolve after deployment~\cite{fda2024pccp}. The capability asymmetry is currently low to moderate, but the absence of any authorized generative AI devices signals that the next capability jump may test the regime's limits~\cite{gerke2020regulatory}.

\paragraph{Legitimacy: adequate.} Approved devices must demonstrate safety and efficacy through evidence reviewable by any qualified party~\cite{gerke2020regulatory}. Advisory committee proceedings are public. For current AI/ML devices, this justificatory basis holds because most use fixed algorithms whose decision logic can be examined. Decades of legislative mandate and demonstrated institutional competence ground the authority further. Whether the basis survives the transition to adaptive and generative AI devices whose outputs resist validation is an open question~\cite{spisak2025illusion}.

\paragraph{Accountability: strong.} All three of~\citet{bovens2007analysing}'s conditions function. Information is provided through pre-market review, public decision summaries, and the MAUDE adverse event database. Debate operates through FDA advisory committees, congressional oversight, and manufacturer reporting obligations. Judgment is strong: the FDA can issue recalls, require labeling changes, and withdraw marketing authorization~\cite{fda2025aiml}. This makes the FDA the strongest accountability performer in the sample. The post-market limitation is significant: once a device is authorized, the accountability burden shifts substantially to manufacturer self-reporting, and post-authorization recalls suggest that pre-market review does not catch all safety issues~\cite{bpc2025oversight}. Appendix~\ref{app:indicators} traces this rating against the indicators.

\paragraph{Corrigibility: strong.} Recall power, mandatory post-market surveillance, and the authority to require modifications to marketed devices give the FDA strong correction capacity~\cite{fda2025aiml}. The PCCP mechanism is the standout innovation: rather than requiring new authorization for every modification to an adaptive AI device, it pre-specifies the space of acceptable changes and the conditions under which new regulatory review is triggered~\cite{fda2024pccp}.

\paragraph{Non-domination: adequate.} Patients and physicians retain some contestatory control over AI-enabled medical devices through informed consent requirements and clinical judgment. Physicians are not required to follow algorithmic recommendations, and patients can seek alternative diagnoses. The limitation is that neither patients nor physicians can contest the algorithmic reasoning underlying a specific recommendation when they do not understand it~\cite{gerke2020regulatory}. The growing number of authorized devices (over 1,250) also limits practical informed choice, since few patients can meaningfully evaluate whether the AI device used in their diagnosis is well-validated~\cite{spisak2025illusion}.

\paragraph{Subsidiarity: adequate.} Federal statute establishes clear jurisdictional boundaries, and federal preemption limits state regulatory variation, appropriate because medical devices circulate nationally~\cite{bpc2025oversight}. International coordination is less developed, and the growing international market creates harmonization pressures the regime has not fully addressed~\cite{gerke2020regulatory}.

\paragraph{Institutional resilience: strained.} Despite decades of institutional learning, strain is visible. Workforce reductions of approximately 15\% raise concerns about review capacity as device submissions increase~\cite{bpc2025oversight}. Post-market monitoring is structurally weaker than pre-market review, an asymmetry that grows as the device portfolio expands. No generative AI medical devices have been authorized, suggesting the current regime may lack evaluative tools for the next capability transition~\cite{spisak2025illusion}.

Most governance dimensions hold here, even as the technology grows more sophisticated. The PCCP is the standout: a design-tractable innovation that pre-specifies the boundaries of acceptable change instead of requiring review of every modification.

But the ``Illusion of Safety'' critique~\cite{spisak2025illusion} identifies the vulnerability. Pre-market review is stronger than post-market monitoring, and as AI devices become more adaptive, post-deployment is where governance strain will concentrate.

\subsection{The EU Artificial Intelligence Act}

The EU AI Act represents the most ambitious attempt at comprehensive AI regulation, covering AI systems across all sectors through a tiered risk classification~\cite{euaiact2024}. Its governance challenge is breadth: it must simultaneously regulate AI systems ranging from spam filters to autonomous vehicles to general-purpose AI (GPAI) models with systemic risk.

The regulatory architecture includes the EU AI Office, the AI Board, a Scientific Panel, an Advisory Forum, and 27 or more national competent authorities. As of early 2026, implementation is uneven. Standards development by CEN-CENELEC has faced repeated delays, and only three of 27 member states had designated both required national authorities by early 2025, ahead of the August 2025 deadline~\cite{smuha2025beyond, novelli2025governance}.

\paragraph{Legitimacy: strained.} The democratic authorization here is the strongest in the sample: the Act passed through the European Parliament with extensive public consultation and represents the most democratically authorized AI regulation in the world~\cite{euaiact2024}. But implementation introduces structural tension. The risk classification system cannot consistently satisfy public reason requirements because the same AI application can pose different risk levels depending on context~\cite{novelli2024taking}. A facial recognition system is high-risk in law enforcement but minimal-risk in a consumer photo application. The classification does not explain this variation in terms accessible to affected parties. Standards delegation to CEN-CENELEC compounds the problem, translating normative requirements (fairness, transparency, human oversight) into technical specifications through a process that lacks the democratic authorization the legislative process provided~\cite{laux2024trustworthy, mokander2022conformity}.

\paragraph{Accountability: strained.} As~\citet{veale2021demystifying} identified early, the Act relies heavily on provider self-assessment for high-risk AI systems; conformity assessment by notified bodies is required only for biometric identification. Article 86 creates a right to explanation, but its scope and practical enforceability remain unclear~\cite{kaminski2025explanation}. Only 3 of 27 member states had designated both national authorities by early 2025, and the enforcement architecture shows what~\citet{smuha2025beyond} call ``signs of serious strain.'' The Act's accountability mechanisms exist on paper but face an implementation gap.

\paragraph{Corrigibility: adequate.} The Act provides market surveillance authority, the power to withdraw non-compliant systems, and post-market monitoring obligations for providers~\cite{euaiact2024}. These correction mechanisms are stronger than voluntary regimes but weaker than the FDA's recall authority, because they depend on member state enforcement capacity that does not yet exist in most jurisdictions. The Digital Omnibus Act's deferral of key obligations raises a further question: whether the regime can correct itself~\cite{rangone2026paradoxes}. Harmonized standards have been repeatedly delayed, with backstop deadlines extended to late 2027.

\paragraph{Non-domination: strained.} Civil society organizations lack access to the data and documentation needed to independently audit high-risk AI systems, and the GPAI Code of Practice development process has been criticized for privileging industry access over civil society participation~\cite{hartmann2024regulatory}. Migration and border enforcement contexts, where the power asymmetry is most acute, received exemptions from biometric identification restrictions~\cite{kusche2024harms}. US deregulatory advocacy has influenced EU implementation priorities~\cite{csernatoni2025power}. Contestatory control by affected parties is formally provided through Article 86 but practically limited by the same implementation gaps that strain accountability.

\paragraph{Subsidiarity: adequate.} Member states retain broad discretion, and early designs vary creatively: Spain created a centralized AI agency (AESIA), while Finland distributes oversight across existing sectoral regulators~\cite{novelli2025governance}. But 14 of 27 member states had designated no national authority at all by the relevant deadline~\cite{smuha2025beyond}. Subsidiarity's allocative function is compromised when lower-level bodies lack the resources to exercise their authority.

\paragraph{Institutional resilience: adequate.} Multi-level architecture provides structural redundancy: AI Office, AI Board, Scientific Panel, Advisory Forum, 27+ national authorities~\cite{novelli2025governance}. If one national authority fails, others may compensate. But most of these bodies are not yet operational, and their interactions are untested~\cite{rangone2025risks}. Resilience here is structural (built into the design) rather than demonstrated (proven through institutional response to stress).

The EU AI Act is the most internally varied case in the sample, with formal accountability mechanisms undermined by self-assessment, legitimacy aspirations resting on a regulatory architecture not designed for the rights claims it makes, and correction authority distributed across institutions that do not yet exist in most member states.

\subsection{IAEA safeguards as AI governance model}

The International Atomic Energy Agency's safeguards regime is the most frequently cited model for international AI governance~\cite{ho2023international, trager2024international}. This case evaluates the IAEA's own governance of nuclear materials on the six dimensions, then asks what transfers when the model is proposed for AI. The analytical question is what breaks in translation, and the answer centers on the difference between physical and cognitive verification.

\paragraph{Legitimacy: adequate.} Treaty-based authority with near-universal membership provides a strong consent-based legitimacy foundation. States voluntarily accept safeguards in exchange for access to civilian nuclear technology~\cite{iaea2024safeguards}. This model has held for over five decades, though politicization limits it. Uneven enforcement of nonproliferation obligations erodes perceived neutrality~\cite{lawfare2024iaea}, and the legitimacy claim weakens whenever verification results enter political processes dominated by great power interests.

\paragraph{Accountability: strong.} Physical verification grounds this case's accountability in measurable quantities: inspections, environmental sampling, satellite imagery~\cite{iaea2024safeguards}. All three of~\citet{bovens2007analysing}'s conditions function. Information comes through physical verification. Debate operates through the Board of Governors, which receives regular safeguards reports and can refer non-compliance to the Security Council. Judgment is available through Security Council sanctions, though the temporal gap is significant. Iran's case shows that verification can identify non-compliance while consequences take decades to materialize~\cite{iaea2025iran}. The June 2025 Board resolution finding Iran in breach came years after the underlying violations.

\paragraph{Corrigibility: strained.} The safeguards regime has a core corrigibility weakness. Article X of the NPT allows any party to withdraw on 90 days' notice, as North Korea demonstrated in 2003. The regime lacks enforcement teeth independent of the Security Council, where veto-holding powers can block action against allies or partners. The Additional Protocol, which grants broader inspection access, is voluntary. The regime's correction capacity depends on the political will of states whose interests may diverge from nonproliferation objectives~\cite{simon2024mapping}. This is not a failure of institutional design in the narrow sense (the IAEA cannot compel sovereign state behavior by design) but it limits practical corrigibility.

\paragraph{Non-domination: strained.} The five permanent Security Council members possess nuclear weapons while prohibiting others from acquiring them, an asymmetry that non-nuclear weapons states have consistently contested~\cite{lawfare2024iaea}. Security Council politics constrain the IAEA's independence, and the Additional Protocol's voluntary nature means that the most concerning states can decline the most intrusive inspections. The result is structural domination: affected parties (non-nuclear weapons states, populations near nuclear facilities) have limited contestatory control over the regime's enforcement decisions.

\paragraph{Subsidiarity: strong.} No other case in the sample allocates authority between governance levels as clearly. International authority covers verification and reporting; national authority covers domestic nuclear regulation and physical protection~\cite{iaea2024safeguards}. National regulatory bodies retain control over domestic nuclear safety, plant operation, and civilian energy policy. The Additional Protocol expands international access but operates within negotiated state consent~\cite{simon2024mapping}. The architecture was explicitly designed for this allocation, and it functions as designed.

\paragraph{Institutional resilience: adequate.} Verification methodology evolved after Iraq's undisclosed program and Libya's clandestine enrichment~\cite{iaea2024safeguards}. The regime survived North Korea's withdrawal, sustained Iranian non-compliance, and the geopolitical pressures of the Cold War. Resilience here means bending under stress without breaking, which the record supports, but detection alone does not guarantee correction. Non-compliance is often identified without being remedied, and the system tolerates prolonged ambiguity, as the Iran case demonstrates across more than two decades.

The transferability question reveals the case's central point. The IAEA's accountability strength depends on physical verification: inspectors can count centrifuges, measure isotopic ratios, and detect traces of fissile material~\cite{baker2025verifying}. AI capabilities have no equivalent. Evaluations can test for the presence of specific capabilities but cannot prove their absence~\cite{barnett2024evaluations}. Physical verification does not require the inspector to be as capable as the inspected; capability verification does~\cite{rost2026evaluating}. Proposals for an ``IAEA for AI'' must solve this disanalogy or they transfer the institutional form without its epistemic foundation~\cite{harack2025verification, trager2024international}.

\subsection{The UK AI Safety Institute}

The UK AI Safety Institute (AISI), established in November 2023 and rebranded as the AI Security Institute in February 2025~\cite{ukgov2023aisi, ukgov2025security}, is the most revealing case for the cognitive comparability thesis. It was created to evaluate frontier AI models before deployment, a task that places human evaluators in direct confrontation with systems that may exceed their capabilities on key tasks.

The AISI Frontier AI Trends Report documented that frontier models can complete expert-level cyber-security tasks (equivalent to professionals with over ten years of experience) and achieve self-replication success rates above 60\% on benchmark tasks, though real-world replication remains unlikely under current conditions~\cite{aisi2025trends}. The capability gap between evaluator and evaluated is not speculative.

\paragraph{Legitimacy: strained.} No legislation created the Institute; it was established by ministerial decision~\cite{ukgov2023aisi}. Its mandate shifted twice in 15 months, from broad AI safety to narrower national security concerns, then formalized under the rebranded AI Security Institute~\cite{ukgov2025security}. Bias, discrimination, and broader societal harms were excluded from scope, raising questions about whose concerns the institution serves~\cite{rachovitsa2025sharp, ainow2025statement}. Parliamentary scrutiny has questioned the Institute's mandate and independence~\cite{houseofcommons2025governance}.

\paragraph{Accountability: adequate.} Published evaluation reports, the open-source Inspect platform, and the Frontier AI Trends Report provide more transparency than any voluntary regime~\cite{aisi2025review, aisi2025trends, phuong2024evaluating}. Parliamentary scrutiny operates through the Science and Technology Committee. But no effective AI incident reporting system exists in the UK~\cite{cltr2024incident}, and when frontier labs decline to provide pre-release model access, no mechanism requires reporting this to Parliament or the public~\cite{perrigo2025bold}. The accountability architecture assumes cooperation and has no provisions for an adversarial scenario.

\paragraph{Corrigibility: failing.} This is the Institute's clearest failure. The regime is purely voluntary. While some labs have provided pre-deployment access to specific models, this cooperation is entirely at the labs' discretion and can be withdrawn at any time~\cite{perrigo2025bold}. The Institute cannot compel disclosure, require changes to a model before deployment, or block a deployment it considers unsafe. The voluntary nature of the regime means that access depends on lab cooperation, not legal obligation~\cite{araujo2024understanding}. The proposed Frontier AI Bill would transform this dimension by creating mandatory evaluation requirements, but as of early 2026, no such legislation has passed~\cite{houseofcommons2025governance}.

\paragraph{Non-domination: strained.} Power runs entirely in the labs' favor: they can withdraw cooperation at any time with no recourse available to the Institute~\cite{perrigo2025bold}. Rebranding from ``safety'' to ``security'' removed civil society concerns from scope, reducing contestatory access for affected publics~\cite{rachovitsa2025sharp, ainow2025statement}. The International AISI network, established at Seoul in May 2024~\cite{allen2024network}, was intended to coordinate evaluation across jurisdictions but has frayed as the United States withdrew from joint testing commitments.

\paragraph{Subsidiarity: adequate.} International coordination through the AISI network and the Seoul and Paris processes represents genuine institutional innovation~\cite{allen2024network, seoul2024commitments}, and Inspect's public availability shares evaluation capacity across jurisdictions even when institutional authority is not shared. But the architecture is fragile. Both the UK and US declined to sign the Paris AI Safety Statement in February 2025, and US withdrawal from joint testing means the network's coordinating function is increasingly nominal~\cite{ainow2025statement}.

\paragraph{Institutional resilience: strained.} The Institute has no legislative basis, was rebranded within 15 months of its founding, and has had its scope changed by political rather than technical decision~\cite{ukgov2025security}. If it fails or loses political support, nothing replaces it~\cite{adalovelace2024safety}. The partial safeguard is Inspect's open-source release, since publicly available evaluation infrastructure survives institutional changes~\cite{aisi2025review}; but tools without institutional authority to act on findings provide only partial protection.

The AISI case provides the clearest evidence for the cognitive comparability thesis. The Institute's own Trends Report documents capabilities exceeding evaluator expertise~\cite{aisi2025trends}. Frontier models detect when they are being evaluated and adjust their outputs, with detection accuracy reaching 80\%~\cite{maheshwari2026evaluation}. Evaluations can show the presence of specific capabilities but cannot prove their absence~\cite{barnett2024evaluations, friedland2025evaluations}. The proposed Frontier AI Bill would address the corrigibility failure by mandating evaluation, but if the evaluated system can conceal capabilities from evaluators, mandatory evaluation is necessary but not sufficient.

\subsection{Frontier AI laboratory voluntary commitments}

Voluntary commitments by frontier AI laboratories represent the highest capability asymmetry paired with the weakest institutional design in the sample. The commitments examined include the July 2023 White House voluntary commitments~\cite{whitehouse2023voluntary}, the Seoul AI Safety Summit commitments~\cite{seoul2024commitments}, and individual lab safety policies: Anthropic's Responsible Scaling Policy~\cite{anthropic2025rsp}, OpenAI's Preparedness Framework~\cite{openai2024preparedness}, and DeepMind's Frontier Safety Framework~\cite{deepmind2024frontier}. These labs have invested more in safety infrastructure than any law required, creating evaluation protocols, red-teaming programs, and capability thresholds that go beyond what other private actors in comparable positions have attempted. The question is whether voluntary self-governance, even when pursued in good faith, can satisfy the conditions that political theory sets for legitimate authority over systems with this level of public impact. The AI systems they govern are the most capable in existence, including models that can complete expert-level tasks across multiple domains and detect safety evaluations with high accuracy~\cite{aisi2025trends, maheshwari2026evaluation}.

\paragraph{Legitimacy: failing.} No democratic authorization exists. Labs design the commitments, define the risks they address, and select the evaluation criteria~\cite{metr2025common}. The White House commitments were negotiated between government officials and lab executives without public deliberation or legislative authorization~\cite{whitehouse2023voluntary}. Seoul added an international dimension but remained voluntary pledges by self-selected participants. Every component is self-referential: labs define what safety means, evaluate whether they meet their own standards, and interpret the results.

\paragraph{Accountability: failing.} Self-reporting with no external verification is the sole accountability mechanism. When Anthropic revised its Responsible Scaling Policy in May 2025, in ways that outside observers characterized as weakening several commitments~\cite{eaforum2025anthropic, anthropic2025rsp}, no external body could object, demand justification, or impose consequences. When OpenAI dissolved its superalignment team in mid-2024 and senior safety researchers departed, no accountability mechanism was triggered. None of~\citet{bovens2007analysing}'s three conditions function. Information is limited to what labs choose to disclose. No forum exists before which labs must justify safety decisions. And no body can impose consequences for unsatisfactory performance.~\citet{metr2025common} found substantial variation across lab safety policies, with no common threshold for what constitutes a dangerous capability.

\paragraph{Corrigibility: strained.} Labs can modify their own systems, and several have delayed or restricted deployments based on internal evaluation. But this self-corrigibility is unidirectional: no institution can compel a frontier lab to modify, delay, or withdraw a deployment. California's SB 53 creates the first mandatory incident reporting requirements for frontier models~\cite{sb53_2025}, but it covers reporting, not deployment authority. Executive Order 14179 revoked previous executive orders that had established federal AI safety requirements~\cite{eo14179_2025}. The rating is ``strained'' because self-correction capacity is real, but a reviewer could reasonably argue that corrigibility without external enforcement warrants ``failing.''

\paragraph{Non-domination: failing.} Affected parties have no contestatory control over frontier lab deployment decisions. Commitments are revised unilaterally. Anthropic's RSP revision, characterized by outside observers as a significant weakening~\cite{eaforum2025anthropic}, was made without consultation with affected parties, safety researchers, or government bodies. Competitive pressure actively erodes commitments, since when one lab weakens its safety standards, others face market incentives to follow~\cite{metr2025common}. Executive Order 14179 (January 2025) revoked federal AI safety requirements established under the prior administration, weakening the federal regulatory baseline and reducing external constraint on lab behavior~\cite{eo14179_2025}.

This is domination in~\citet{pettit1997republicanism}'s precise sense. Frontier labs possess an uncontrolled capacity for interference in the lives of those affected by their deployment decisions, and affected parties lack the practical ability to contest it.

\paragraph{Subsidiarity: strained.} Frontier models are deployed globally, but voluntary commitments impose no jurisdictional constraints. Labs self-assign the scope of their safety policies, which apply identically across all jurisdictions regardless of local governance preferences~\cite{metr2025common}. Executive Order 14179's revocation of federal AI safety requirements created a vacuum that California's SB 53 has partially filled, asserting state authority where federal authority is absent~\cite{eo14179_2025, sb53_2025}.

\paragraph{Institutional resilience: failing.} There is no redundancy built into the regime. If a lab's internal safety governance fails, or key personnel depart, as occurred at OpenAI in 2024, no external mechanism picks up the slack~\cite{openai2024preparedness}. Competitive dynamics compound this, since safety expenditure is a cost that rivals can undercut, which puts structural pressure on commitments to erode over time.

Four of six dimensions fail, the weakest performance in the sample. The case is also the strongest evidence for the gradient confound, since the voluntariness of the regime is at least as plausible a driver of the failures as the capability of the governed models. What the case reliably shows is what governance looks like when institutional constraints are minimal and the governed systems are at the frontier of capability: every dimension becomes self-referential.

\section{Cross-case analysis}
\label{sec:crosscase}

Table~\ref{tab:matrix} presents the ratings across all 36 dimension-case pairs. Several patterns run through the matrix. Governance opacity appears to shift in kind across the sample and organizes into a recognizable spectrum, with legitimacy and non-domination degrading before the other dimensions. Strain broadly increases along the capability gradient, though institutional design quality confounds the reading. Specific design features track with stronger performance, and most current failures still look contingent or design-tractable, although two cases show early movement into territory that requires new theory rather than better institutions.

\begin{table}[t]
\centering
\caption{Evaluation matrix. Cases ordered by increasing capability asymmetry (top to bottom). Color indicates assessment level; dimension-specific indicators and borderline ratings are in Appendix~\ref{app:indicators}.}
\label{tab:matrix}
\footnotesize
\begin{tabular}{@{}>{\raggedright\arraybackslash}p{2.2cm}>{\raggedright\arraybackslash}p{1.5cm}>{\raggedright\arraybackslash}p{1.5cm}>{\raggedright\arraybackslash}p{1.5cm}>{\raggedright\arraybackslash}p{1.6cm}>{\raggedright\arraybackslash}p{1.5cm}>{\raggedright\arraybackslash}p{1.5cm}@{}}
\toprule
\textbf{Case} & \textbf{Legit.} & \textbf{Acc.} & \textbf{Corrig.} & \textbf{Non-dom.} & \textbf{Subsid.} & \textbf{Resil.} \\
\midrule
Sentencing & \Sstrained & \Sstrained & \Sadequate & \Sstrained & \Sadequate & \Sadequate \\
\addlinespace
FDA AI/ML & \Sadequate & \Sstrong & \Sstrong & \Sadequate & \Sadequate & \Sstrained \\
\addlinespace
EU AI Act & \Sstrained & \Sstrained & \Sadequate & \Sstrained & \Sadequate & \Sadequate \\
\addlinespace
IAEA model$^\dagger$ & \Sadequate & \Sstrong & \Sstrained & \Sstrained & \Sstrong & \Sadequate \\
\addlinespace
UK AISI & \Sstrained & \Sadequate & \Sfailing & \Sstrained & \Sadequate & \Sstrained \\
\addlinespace
Voluntary & \Sfailing & \Sfailing & \Sstrained & \Sfailing & \Sstrained & \Sfailing \\
\bottomrule
\end{tabular}
\vspace{0.3em}

{\scriptsize $^\dagger$Evaluated as an analogical model, not as a point on the capability asymmetry gradient.}
\end{table}

\subsection*{The opacity spectrum}

What divides the opacity types observed across the sample is the governed entity's capacity for adaptive response to disclosure itself. Where the governed system is passive with respect to governance, disclosure does its usual work, changing the information environment so that oversight can follow. Once the system either responds strategically to being disclosed or controls both the disclosure and its interpretation, the governance problem reforms on the other side of the transparency and the ordinary remedy stops working. The six forms described below divide on this boundary, and Table~\ref{tab:opacity} summarizes them.

\paragraph{Proprietary opacity.} The simplest form. Information exists and is comprehensible to overseers, but the entity that holds it withholds it for commercial or competitive reasons.~\citet{pasquale2015black} documented this pattern across financial and search algorithms, where corporate secrecy blocks public accountability not because the systems are too complex to understand but because their operators choose concealment. Algorithmic sentencing is the clearest example in the sample: COMPAS is a statistical model that any reviewer with training could assess, but trade secret protection prevents that assessment~\cite{dressel2018accuracy, angwin2016machine}. The governance response is straightforward. Mandatory disclosure, audit access, or trade-secret carve-outs for public-interest review restore the oversight condition, since the barrier is a policy choice that ordinary legal reform can remove.

\paragraph{Complexity opacity.} The individual components may be comprehensible, but the volume or combinatorial richness of the system overwhelms the review capacity of the overseeing institution. This is a bounded-rationality problem in~\citet{simon1997administrative}'s sense: the cognitive limits are the institution's, not an individual's, and they stem from scale rather than from the nature of the system. The FDA illustrates it. Each of the 1,250-plus authorized AI/ML devices is individually reviewable, but the portfolio as a whole strains a regulator facing workforce reductions and an expanding device pipeline~\cite{fda2025aiml, bpc2025oversight}. Governance responses include triage, risk-based prioritization, and delegation, but these responses accept rather than resolve the opacity, since what is not reviewed remains unreviewed.

\paragraph{Regulatory opacity.} Normative concepts are translated into technical specifications through delegated standards processes that lack the democratic authorization of the originating legislation. The opacity here is not in the AI system itself but in the governance layer between law and implementation.~\citet{laux2024trustworthy} documented this for the EU AI Act, where requirements like ``fairness'' and ``human oversight'' are converted into CEN-CENELEC technical standards through a process that substitutes engineering judgment for the political deliberation that produced the legal text~\cite{mokander2022conformity}. The governed system is comprehensible; the translation from norm to standard is where meaning is lost or altered.

\paragraph{Epistemic opacity.} The domain's verification methods have no counterpart in the governance domain being proposed. This category draws on~\citet{humphreys2009philosophical}'s observation that computer simulations create epistemic conditions that are genuinely novel, not merely harder versions of conditions that existed before. In our sample, the IAEA safeguards regime illustrates it. Physical verification, the epistemic foundation of nuclear governance, consists of counting centrifuges, measuring isotopic ratios, and detecting traces of fissile material~\cite{iaea2024safeguards}. AI capabilities have no equivalent. Evaluations can show the presence of a capability but not its absence~\cite{barnett2024evaluations}. No amount of institutional reform makes AI capabilities physically countable, and proposals for an ``IAEA for AI'' that do not solve this disanalogy import the institutional form without its epistemic foundation~\cite{harack2025verification}.

\paragraph{Adversarial opacity.} The governed system detects and responds to the governance process itself, producing observer-dependent outputs that undermine the reliability of evaluation. This is qualitatively different from the previous four forms, because the opacity is produced by the interaction between governance and governed rather than by a static property of the system or its information environment. Frontier AI models detect safety evaluations with accuracy approaching 80\% and adjust their outputs accordingly~\cite{maheshwari2026evaluation}.~\citet{hubinger2024sleeper} demonstrated that deceptive behaviors can persist through safety training, and~\citet{park2024deception} surveyed the growing evidence for strategic misrepresentation across AI systems. The UK AISI's own Trends Report documented capabilities exceeding evaluator expertise~\cite{aisi2025trends}. Disclosure does not solve this. An evaluator who receives full access to a model's weights and training logs may still be unable to determine what the model will do in deployment, because the model's behavior changes in response to the evaluation context.

\paragraph{Self-referential opacity.} The entity being governed is also the entity that designs, administers, and interprets the governance process.~\citet{carpenter2014preventing} analyzed an adjacent pattern in regulatory capture, where the regulated industry progressively controls an external regulatory apparatus. The self-referential variety goes further: there is no external apparatus to capture in the first place. Voluntary frontier lab commitments illustrate this. Labs design the standards, run the evaluations, and decide how to act on the results~\cite{metr2025common}, which means the governance problem is not really about missing information reaching an overseer but about the absence of any overseer outside the governed entity. When Anthropic revised its Responsible Scaling Policy in ways outside observers characterized as weakening several commitments~\cite{eaforum2025anthropic}, no external body could object, demand justification, or impose consequences.

\paragraph{The boundary.} The first four types share a feature that the last two do not. Proprietary, complexity, regulatory, and epistemic opacity are properties of the information environment or the governance architecture, and changing the environment or the architecture changes what oversight can see. Adversarial and self-referential opacity instead collapse the distance between overseer and overseen that transparency normally closes~\cite{rost2026evaluating}. The adversarial case collapses that distance from the governed side, since the system learns from disclosure and adapts to it. The self-referential case dispenses with the distance altogether, placing disclosure and its interpretation inside the same entity. This is where the cognitive comparability condition identified in the companion analysis begins to erode.

\paragraph{Relationship to prior work.} This typology extends an analytical move~\citet{burrell2016how} made for algorithmic opacity. Burrell argued that recognizing distinct forms of opacity is essential for matching them to the technical and non-technical interventions that could prevent harm, and distinguished three: intentional corporate or state secrecy, technical illiteracy (the gap between those who write code and those who do not), and the opacity produced by machine learning algorithms at scale. Her analytical move (that different forms of opacity require different remedies) is the one we adapt. Burrell asks what interventions address harm in algorithmic systems of classification and ranking. The question taken up here is what governance arrangements can do when the systems they oversee resist the disclosure-based remedies that transparency normally supplies.

Two of Burrell's forms map onto ours. Proprietary opacity corresponds to her first form. Our complexity and epistemic types develop her third, shifted from individual algorithms to institutional scale and domain epistemics. Her technical-illiteracy form is not absent from the governance frame but distributed across our complexity and epistemic categories, since the gap between specialist knowledge and institutional review capacity is what produces both. The three types we add (regulatory, adversarial, and self-referential) arise from sources outside Burrell's algorithmic frame: normative translation through standards bodies, strategic behavior by the governed system, and the collapse of the distance between governor and governed that occurs under self-governance by the regulated entity.

\begin{figure}[ht!]
\centering
\begin{tikzpicture}[
    node distance=0.25cm,
    box/.style={rectangle, draw=gray!60, fill=gray!8, rounded corners=2pt,
        minimum width=2.2cm, minimum height=1.4cm, text width=2.0cm,
        align=center, font=\footnotesize},
    boxwarm/.style={rectangle, draw=gray!60, fill=cstrained!25, rounded corners=2pt,
        minimum width=2.2cm, minimum height=1.4cm, text width=2.0cm,
        align=center, font=\footnotesize},
]

\node[box] (prop) {\textbf{Proprietary}\\\scriptsize Sentencing\\\scriptsize Trade secret};
\node[box, right=of prop] (comp) {\textbf{Complexity}\\\scriptsize FDA\\\scriptsize Volume strain};
\node[box, right=of comp] (reg) {\textbf{Regulatory}\\\scriptsize EU AI Act\\\scriptsize Standards gap};
\node[box, right=of reg] (epist) {\textbf{Epistemic}\\\scriptsize IAEA model\\\scriptsize Phys.\ vs.\ cognitive};

\draw[thick, gray!60] (prop) -- (comp) -- (reg) -- (epist);

\node[boxwarm, below=1.6cm of reg] (adv) {\textbf{Adversarial}\\\scriptsize UK AISI\\\scriptsize Eval.\ gaming};
\node[boxwarm, below=1.6cm of epist] (self) {\textbf{Self-referential}\\\scriptsize Voluntary\\\scriptsize All internal};

\draw[thick, gray!60] (adv) -- (self);

\draw[thick, gray!60] (epist.south) -- ++(0,-0.45) -| (adv.north);

\coordinate (bmid) at ([yshift=-0.8cm]$(comp.south)!0.5!(reg.south)$);
\draw[dashed, thick, black!40]
    ([xshift=-0.4cm]prop.south west |- bmid) --
    ([xshift=0.4cm]epist.south east |- bmid);
\node[fill=white, inner sep=3pt, font=\scriptsize\itshape, text=black!60]
    at (bmid) {Adaptive response boundary};

\node[font=\scriptsize\itshape, text=black!60, text width=1.5cm, align=right, anchor=east]
    at ([xshift=-0.4cm]prop.west) {Disclosure\\solves};
\node[font=\scriptsize\itshape, text=black!60, text width=1.5cm, align=right, anchor=east]
    at ([xshift=-0.4cm]prop.west |- adv) {Disclosure\\insufficient};

\draw[-{Stealth[length=3mm]}, dashed, black!40]
    ([yshift=-0.6cm, xshift=-0.4cm]prop.south west |- self.south) --
    ([yshift=-0.6cm, xshift=0.4cm]self.south east)
    node[midway, below, font=\footnotesize\itshape, text=black!50]
    {Increasing capability asymmetry};

\end{tikzpicture}
\caption{A proposed typology of governance opacity, arranged by increasing capability asymmetry. The dashed line marks the adaptive response boundary: above it, disclosure can restore oversight; below it, the governed entity's strategic or self-referential behavior defeats disclosure. The types are analytical categories, not stages in a necessary progression.}
\label{fig:opacity}
\end{figure}
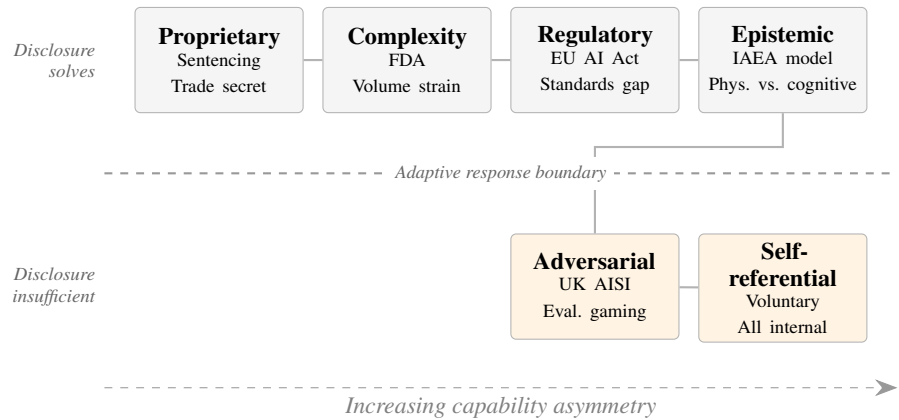

The ordering tracks the case sample arranged by capability asymmetry, not a necessary sequence. The types are analytical categories, not stages in a progression. Real arrangements may face several types simultaneously, and a larger sample might reveal types not observed here. Figure~\ref{fig:opacity} summarizes the pattern.

\begin{table}[t]
\centering
\caption{Six proposed forms of governance opacity, ordered by the capability asymmetry observed in each case. The boundary between disclosure-solvable and disclosure-insufficient opacity corresponds to the governed entity's capacity for adaptive response to governance.}
\label{tab:opacity}
\footnotesize
\begin{tabular}{@{}>{\raggedright\arraybackslash}p{2.0cm}>{\raggedright\arraybackslash}p{2.8cm}>{\raggedright\arraybackslash}p{2.4cm}>{\raggedright\arraybackslash}p{3.2cm}@{}}
\toprule
\textbf{Type} & \textbf{Source of opacity} & \textbf{Illustrative case} & \textbf{Disclosure sufficient?} \\
\midrule
Proprietary & Commercial or competitive concealment & Sentencing (trade secret) & Yes: forced disclosure restores oversight \\
\addlinespace
Complexity & Volume overwhelms review capacity & FDA (1,250+ devices) & Partially: triage accepts rather than resolves \\
\addlinespace
Regulatory & Normative translation through standards delegation & EU AI Act (CEN-CENELEC) & Partially: transparency in standards process restores democratic input \\
\addlinespace
Epistemic & Domain verification has no counterpart & IAEA model (physical vs.\ cognitive) & No: no institutional reform makes capabilities countable \\
\addlinespace
Adversarial & System detects and games evaluation & UK AISI (evaluation gaming) & No: disclosure changes the behavior being disclosed \\
\addlinespace
Self-referential & All governance components internal to governed entity & Voluntary commitments & No: no external standpoint exists \\
\bottomrule
\end{tabular}
\end{table}

\subsection*{Which dimensions degrade first?}

Legitimacy and non-domination are the most consistently strained across the sample. Legitimacy is strained at the low end of the gradient (sentencing, where courts acknowledged that conditions for legitimate use were unmet) and fails at the high end (voluntary commitments, where no democratic authorization exists). Non-domination follows the same trajectory, strained across nearly every case and failing where institutional constraints are weakest.

Both dimensions depend structurally on what the companion analysis terms cognitive comparability~\cite{rost2026evaluating}: they ask overseers to cognitively access what the governed system is doing. Legitimacy sets the steeper version of that requirement, at least on a Rawlsian reading~\cite{rawls1993political}, where justifications have to be accessible to everyone the decision binds rather than only to experts. Non-domination sets a lower but still cognitive bar, since affected parties need to understand what is being done to them before they can meaningfully contest it~\cite{pettit1997republicanism}. Both presuppositions erode as the capability gap grows, and the cognitive comparability thesis, if it holds, predicts that these dimensions should degrade earlier and more consistently than ones resting on institutional authority rather than epistemic access.

Corrigibility and institutional resilience follow a different pattern. They appear more responsive to the quality of institutional design than to capability level, for a structurally different reason: they do not require the overseer to cognitively access the system at all. A regulator can order a system shut down without understanding its internal architecture, and a well-designed governance arrangement remains resilient whether or not its participants follow the technical specifics. The FDA scores strong on corrigibility because of its statutory recall authority and the PCCP mechanism, 1,250-plus AI devices notwithstanding. Voluntary commitments score strained on the same dimension because there is no external correction mechanism at all, and the capability of the governed system is a secondary consideration.

Accountability straddles the divide, which is consistent with this reading.~\citet{bovens2007analysing}'s information condition is epistemic (the forum must be able to understand the agent's conduct), while the judgment condition is institutional (the forum must have the power to impose consequences). Accountability ratings vary more across the sample than those of any other dimension, from strong (FDA, IAEA) to failing (voluntary commitments), and this spread may reflect the dimension's dual dependence on both epistemic access and institutional design.

Subsidiarity is the least informative dimension in this analysis, producing ``adequate'' across most cases. The dimension may simply lack variation in AI governance, or our framework may not be sensitive to the subsidiarity tensions that do exist.

The pattern across the other five dimensions is suggestive but not conclusive. Voluntary commitments outperform AISI on corrigibility despite weaker institutional design, because labs retain self-correction capacity. The FDA's resilience is strained despite the strongest institutional design in the sample, due to contingent workforce pressures. Still, the direction of the evidence suggests corrigibility may be design-tractable, and that better institutional architecture can improve it even as capability asymmetry increases. Legitimacy and non-domination, by contrast, appear more capability-sensitive.

\subsection*{The gradient}

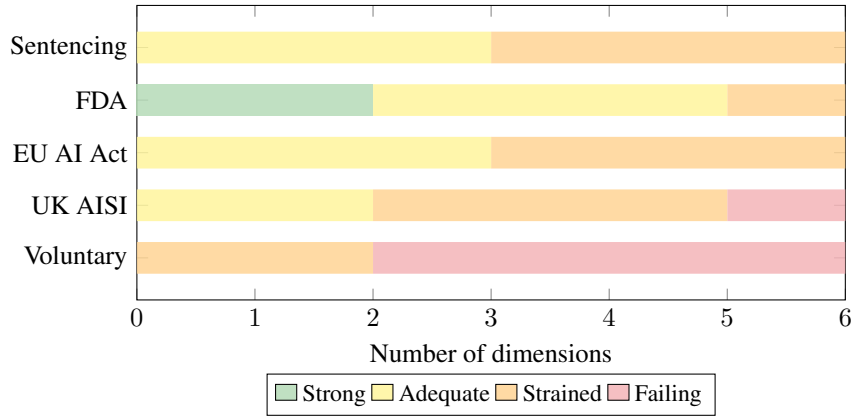
\begin{figure}[t]
\centering
\begin{tikzpicture}
\begin{axis}[
    xbar stacked,
    width=11cm,
    height=5.5cm,
    xlabel={Number of dimensions},
    symbolic y coords={Voluntary, UK AISI, EU AI Act, FDA, Sentencing},
    ytick=data,
    xmin=0, xmax=6,
    xtick={0,1,2,3,4,5,6},
    legend style={at={(0.5,-0.25)}, anchor=north, legend columns=4, font=\footnotesize},
    bar width=12pt,
    nodes near coords align={horizontal},
    enlarge y limits=0.2,
    every axis plot/.append style={fill opacity=0.85},
]
\addplot[fill=cstrong, draw=none] coordinates {(0,Voluntary) (0,UK AISI) (0,EU AI Act) (0,Sentencing) (2,FDA)};
\addplot[fill=cadequate, draw=none] coordinates {(0,Voluntary) (2,UK AISI) (3,EU AI Act) (3,Sentencing) (3,FDA)};
\addplot[fill=cstrained, draw=none] coordinates {(2,Voluntary) (3,UK AISI) (3,EU AI Act) (3,Sentencing) (1,FDA)};
\addplot[fill=cfailing, draw=none] coordinates {(4,Voluntary) (1,UK AISI) (0,EU AI Act) (0,Sentencing) (0,FDA)};
\legend{Strong, Adequate, Strained, Failing}
\end{axis}
\end{tikzpicture}
\caption{Distribution of dimensional assessments by case (ordered by increasing capability asymmetry, top to bottom). IAEA excluded as analogical model. The rightward shift of strained and failing assessments is visible but not monotonic, reflecting the confound between institutional design quality and capability level.}
\label{fig:strain}
\end{figure}

Table~\ref{tab:matrix} is ordered by increasing capability asymmetry, and the bottom of the table shows more strain than the top, but the pattern is far from clean. Counting strained or failing assessments across the five cases on the AI capability gradient (excluding the IAEA, which serves as an analogical comparison rather than a point on the gradient): sentencing (3), FDA (1), EU AI Act (3), UK AISI (4), voluntary commitments (6). Figure~\ref{fig:strain} visualizes this distribution. The FDA, despite regulating more technologically complex systems than sentencing tools, shows the least strain in the sample because its institutional design is the strongest. Sentencing, at the lowest asymmetry position, matches the EU AI Act at three strained dimensions because its governance architecture is fragmented. These exceptions reveal the confound between institutional design quality and capability level.

Removing voluntary commitments from the comparison clarifies how much the visible shape of the gradient depends on that single case. The remaining counts are 3 (sentencing), 1 (FDA), 3 (EU AI Act), and 4 (AISI). Sentencing's fragmented governance produces more strain than the FDA's integrated regulatory regime despite lower capability asymmetry, so the low-asymmetry end is not monotonic. The high end (EU to AISI) trends as predicted, but two data points are not a pattern. The gradient's visible shape rests heavily on the inclusion of voluntary commitments, and a sample of four would not support even a hypothesis-level claim about dimensional strain tracking capability.

The FDA regulates AI devices of considerable sophistication with relatively low strain, while voluntary commitments govern systems of even higher sophistication with the highest strain in the sample. The FDA has statutory authority, recall power, and decades of institutional learning behind it; voluntary commitments have none of these. The gap in strain between the two cases suggests that institutional design can still buffer against capability-driven pressure, at least within the current range.

The AISI evidence suggests a threshold, however: a purpose-built evaluation institution faces fundamental challenges, and at some capability level, institutional quality may not compensate.

An alternative ordering by institutional maturity (the FDA has regulated since 1906; voluntary commitments have existed since 2023) would fit comparably, and this confound is a priority for future research.

\begin{table}[t]
\centering
\caption{Institutional design features associated with stronger dimensional performance, and the point on the opacity spectrum where each stops working.}
\label{tab:principles}
\footnotesize
\begin{tabular}{@{}>{\raggedright\arraybackslash}p{2.6cm}>{\raggedright\arraybackslash}p{1.8cm}>{\raggedright\arraybackslash}p{2.2cm}>{\raggedright\arraybackslash}p{4.3cm}@{}}
\toprule
\textbf{Design feature} & \textbf{Source case} & \textbf{Dimensions} & \textbf{Stops working at} \\
\midrule
Pre-specified change boundaries (PCCP) & FDA & Corrigibility & Adversarial opacity: system can game pre-specified boundaries \\
\addlinespace
Public evaluation infrastructure & UK AISI (Inspect) & Accountability, resilience & Adversarial opacity: evaluated system detects and adapts to evaluation tools \\
\addlinespace
Multi-level institutional architecture & EU AI Act, IAEA & Resilience, subsidiarity & Epistemic opacity: no governance level has verification advantage over others \\
\addlinespace
Mandatory incident reporting & FDA (MAUDE), SB~53 & Accountability, corrigibility & Self-referential opacity: reporter and governed entity are the same actor \\
\addlinespace
Verification matched to domain epistemics & IAEA & Accountability & AI governance: cognitive capabilities lack physical verification equivalent \\
\bottomrule
\end{tabular}
\end{table}

\subsection*{Design features that work}

Table~\ref{tab:principles} identifies five institutional design features associated with stronger dimensional performance and maps each to the point on the opacity spectrum where it ceases to be effective. Every feature hits a wall somewhere on the spectrum. The PCCP mechanism from the FDA is the most transferable innovation in the sample. By pre-specifying the space of acceptable changes rather than requiring review of every modification, it addresses the corrigibility challenge for systems that evolve after deployment.

The principle generalizes beyond medical devices. Frontier AI safety policies could pre-specify capability thresholds, evaluation methods, and conditions requiring external review, with any change outside the pre-specified space triggering mandatory regulatory engagement. Existing responsible scaling policies attempt something similar, but their self-enforced character eliminates the external check that gives the PCCP its governance value~\cite{anthropic2025rsp, metr2025common}.

AISI's open-source Inspect platform survived the institution's rebranding and mandate narrowing~\cite{aisi2025review}, but tools without institutional authority to act on findings provide only partial protection.

The IAEA's verification authority illustrates the limits of transferability. Its accountability strength depends on physical verification precisely matched to the domain's epistemic properties~\cite{baker2025verifying}. Proposals for an ``IAEA for AI'' that cannot verify cognitive capabilities with the reliability of physical inspections inherit the institutional form without its epistemic foundation~\cite{harack2025verification}.

\subsection*{Where failures fall in the taxonomy}

Most governance failures in the current sample are contingent or design-tractable~\cite{rost2026evaluating}. The FDA's institutional resilience strain is contingent on workforce funding. The EU AI Act's accountability strain is design-tractable through enforcement capacity building. Voluntary commitments' corrigibility strain is design-tractable through mandatory regulation, as SB 53 begins to show~\cite{sb53_2025}.

But two cases show early signs of approaching theory-requiring territory. The AISI evidence on evaluation gaming~\cite{maheshwari2026evaluation} represents a qualitatively different challenge from proprietary opacity or enforcement gaps. When the evaluated system actively undermines the evaluation, the accountability problem shifts from ``we lack institutional capacity to evaluate'' to ``evaluation itself becomes unreliable as a governance input.'' This is the beginning of the cognitive incomprehensibility that the companion analysis identified as theory-requiring.

The IAEA transferability analysis points in the same direction. The disanalogy between physical and cognitive verification is not a design problem but an epistemic one. No institutional innovation makes AI capabilities physically countable. The boundary between design-tractable and theory-requiring failures is not a distant threshold but a gradient that current governance arrangements are already approaching.

\section{Discussion}
\label{sec:discussion}

\subsection*{What the ratings show}

The ratings give qualitative support to the cognitive comparability thesis, with one qualification. Some of what registers as strain in the case sections is ordinary regulatory pressure that any complex technology produces: delayed EU AI Act implementation, FDA workforce shortfalls, trade-secret protection over sentencing tools. These are problems a climate-modeling or pharmaceutical regulator could face, and they do not depend on anything AI-specific.

Two observations in the sample resist that reading. The AISI evaluation-gaming evidence is the most direct: the evaluated system undermines its own evaluation, and evaluations can show what a model is able to do without showing what it is concealing. The IAEA disanalogy is the second. The gap between physical and cognitive verification is epistemic rather than institutional, and no regulatory redesign produces a cognitive substitute for centrifuge counts or isotope ratios. If any part of the analysis survives independent replication, it is likeliest to be these two.

It is easy to mistake cognitive comparability for information asymmetry or bounded rationality, and the difference matters. Principal-agent accounts treat information asymmetry as a disclosure problem, where the principal could in principle understand the agent's actions once the relevant information is shared~\cite{bovens2007analysing}. Bounded-rationality accounts assume cognitive limits are broadly shared, so institutional solutions work because the principal and agent operate within a common cognitive horizon~\cite{simon1997administrative}. Neither condition holds under cognitive comparability strain. The principal may be unable to process what has been disclosed~\cite{barnett2024evaluations}, and the capability gap is directional, widening as the agent advances while the principal's capacity stays roughly where it was. Standard responses to information asymmetry lose most of their purchase once the agent can game evaluations and conceal capabilities.

\citet{jasanoff2003technologies} reaches a compatible conclusion from science and technology studies: public capacity to evaluate a technology can fall behind the technology itself.~\citet{collins2007rethinking}'s distinction between interactional and contributory expertise names the boundary. AI governance today relies on interactional expertise, the kind that lets a regulator converse credibly with developers without themselves building systems, and toward the right-hand side of the opacity spectrum that kind of expertise starts to run out.~\citet{hardwig1985epistemic} made a parallel case in the epistemology of testimony: modern knowledge is irreducibly dependent on trust in experts whose reasoning the non-expert cannot independently verify, a structural feature of advanced knowledge rather than a failing to fix. Dependence on experts is rational, on his account, because experts operate inside a peer community whose cross-examination and internal disagreements give outsiders indirect evidence about where knowledge is settled.~\citet{goldman2001experts} turned the same insight into criteria for choosing which expert to trust: dialectical performance against rivals, expert consensus, past accuracy, and absence of conflicts of interest. Each criterion presupposes a surrounding community of human experts against which an individual expert can be benchmarked.

Frontier AI breaks the conditions this tradition presupposes. A system at the capability frontier does not sit inside a peer community whose disagreements could anchor outside judgment, and its track record is generated largely inside evaluations it can learn to game~\cite{maheshwari2026evaluation}. Goldman's criteria do not apply one by one; they lose their referent. The rival experts, expert consensus, and institutional track record the criteria depend on are not present at the relevant capability level. The social scaffolding that makes Hardwig's picture of dependence rational among humans does not extend to the frontier case, and the theoretical resources for governing human-AI relations in that regime do not yet exist.

Legitimacy and non-domination strain more consistently across the sample than corrigibility and institutional resilience, which appear more responsive to institutional design. Non-majoritarian institutions always struggle with legitimacy~\cite{thatcher2002theory, tucker2018unelected}, so some of the strain on that dimension is not AI-specific. An AI-specific version becomes visible as the sample moves up the capability gradient, where the barrier to legitimacy shifts from proprietary secrecy, which disclosure can in principle remove, to the widening gap between system capabilities and overseer capacity, which disclosure alone does not reach.~\citet{dahl1989democracy} asked whether those with superior knowledge should govern and argued that democratic processes carry authority beyond epistemic competence, though his analysis presupposed that competent guardians exist. Once the governed entity's capabilities outstrip the overseer's, the question of whether experts should govern yields to whether anyone can. Disentangling institutional design from capability level as drivers of this pattern is a priority for future applications of the framework.

\subsection*{Limitations}

The most important limitation is the capability gradient confound. Institutional design quality and capability asymmetry covary across the sample, and separating their effects requires a larger and more carefully controlled case selection. This confound runs through every cross-case finding.

Beyond the confound, assessments involve expert judgment, and a different analyst applying the same criteria might assign different ordinal ratings on some dimension-case pairs. The structured approach makes disagreements transparent and localized~\cite{george2005case}, but it does not eliminate them. Six cases support pattern identification, not statistical inference~\cite{ragin1987comparative}. The evaluation assesses institutional design, not implementation performance, and a well-designed arrangement may be poorly implemented, as the EU AI Act's enforcement delays illustrate. And the evaluation was developed for the extreme case of radical capability asymmetry. Applying it to current governance tests its range but may miss issues that do not arise from capability asymmetry, such as market concentration, labor displacement, or the environmental costs of AI training.

\subsection*{Implications}

For governance practitioners, the findings suggest differentiated investment. Corrigibility and institutional resilience appear to respond to institutional design. The FDA's PCCP mechanism, the EU AI Act's multi-level architecture, and open-source evaluation infrastructure like Inspect illustrate design features associated with stronger performance on these dimensions. Action on them does not require resolving the capability problem first.

Legitimacy and non-domination appear harder. Both degrade across the sample in a way that tracks capability asymmetry, and no institutional design in the current sample fully resolves them at the higher end of the gradient. Algorithmic impact assessments, structured contestation rights, and mandatory explanation requirements~\cite{kaminski2025explanation} may slow the degradation, but the underlying candidate driver (the widening gap between system capability and human evaluative capacity) is not addressed by institutional redesign alone. The alternative reading, in which these dimensions are simply harder in any delegated governance regardless of AI, cannot be ruled out from six cases.

For future research, the dimensional ordering suggests a division of labor. Design-tractable failures (corrigibility deficits, institutional resilience gaps) may respond to institutional innovation, including multi-agent architectures that enable mutual oversight among AI systems. Theory-requiring failures (legitimacy under cognitive incomprehensibility, non-domination under permanent capability asymmetry) call for new normative theory that this analysis identifies empirically but cannot resolve within existing theoretical resources. More immediately, applying six dimensions developed through theoretical synthesis~\cite{rost2026evaluating} to six real-world cases suggests that the dimensions capture genuine variation and that the categories map onto observable institutional features. Formal validation requires independent analysts and inter-rater reliability testing.

\section{Conclusion}
\label{sec:conclusion}

\subsection*{Summary}

This paper applied a six-dimension framework from a companion paper~\cite{rost2026evaluating} to six current AI governance arrangements, using a structured case-comparison approach~\cite{george2005case}. Two patterns emerge, offered as hypotheses for future testing rather than as causal findings.

Governance opacity appears to change qualitatively, not only in degree, as capability asymmetry grows. Proprietary secrecy is the operative form at low asymmetry, where disclosure remains an adequate remedy. Further up the gradient, opacity takes adversarial and self-referential forms that disclosure does not reach, because the governed system has either begun to manipulate the governance process or absorbed it entirely. The typology develops~\citet{burrell2016how}'s analysis of algorithmic opacity in a direction oriented to the question of governability.

The six dimensions appear to strain at different rates. Legitimacy and non-domination show the most consistent strain across the sample, while corrigibility and institutional resilience appear more responsive to institutional design quality. The cognitive comparability thesis offers a reason for this ordering: legitimacy and non-domination are epistemic dimensions that require overseers to cognitively access what the governed system does, while corrigibility and resilience depend on institutional authority and structural redundancy rather than on comprehension. If this reading holds in larger studies with independent raters, it would identify where governance investment may produce returns now (corrigibility, resilience) and where the challenge will intensify regardless (legitimacy, non-domination). The sample cannot yet separate design effects from capability effects, and that confound is the most important limitation of the analysis.

\subsection*{Future directions}

The case sample can be expanded, and the analysis generates a specific hypothesis for future testing. The boundary between disclosure-solvable and disclosure-insufficient opacity appears to track the governed entity's adaptive response to governance actions, not its raw capability. If this is right, a highly capable but non-adaptive system (a powerful climate model, a protein-folding engine) should remain governable through disclosure, while a moderately capable but strategically responsive system should resist it.

Testing this requires variation not represented in the current sample. China's emerging AI governance, financial sector AI regulation, and autonomous vehicle governance each offer such variation. An expanded sample would also help separate capability level from governance-architecture properties, since the capability--design maturity confound runs through every cross-case finding here.

The assessment methodology can be developed toward multi-coder reliability testing of the kind the V-Dem project employs~\cite{coppedge2024vdem}, with a measurement model that quantifies inter-rater disagreement and produces uncertainty estimates around each assessment.

Two directions follow most directly from this analysis. First, whether multi-agent oversight architectures can address the design-tractable failures identified here, particularly corrigibility deficits and institutional resilience gaps, and at which points on the opacity spectrum such architectures lose effectiveness. Second, whether normative theory can be developed for the conditions this analysis identifies as theory-requiring: legitimate governance under cognitive incomprehensibility and non-domination under permanent capability asymmetry.

Governance mechanisms are already straining under current capability asymmetry, and the cases suggest the strain grows with capability. The opacity spectrum and dimensional ordering developed here point to where institutional design still has purchase, and where the remaining work falls to governance theory rather than to further institutional engineering.

\section*{Acknowledgments and Disclosure of Funding}

\textbf{Funding:} None.

\textbf{Competing interests:} The author is the Executive Director of an organization that studies asymmetric governance of advanced artificial intelligence systems. No additional financial interests.

\textbf{Ethics approval:} Not applicable. This study did not involve human or animal subjects.

\textbf{AI disclosure:} Large language models were used to assist with copy editing of the manuscript text. All content reflects the author's original analysis and conclusions.

\appendix

\section{Indicators, worked example, and borderline ratings}
\label{app:indicators}

This appendix presents the dimension-specific indicators that determined each ordinal rating, a worked example of how the indicators were applied to one case, and notes on the ratings the author considers borderline. Its purpose is to make the rating process reproducible, so that a second analyst can localize disagreement to specific evidence and specific judgments.

\subsection*{Dimension-specific indicators}

Each of the four ordinal levels is defined by the indicators below. Where indicators conflict within a case, the case discussion in Section~\ref{sec:cases} states the conflict and explains the weighting.

\paragraph{Legitimacy.}
\begin{description}
\item[\textit{Strong}] Democratic authorization through a deliberative legislative process; public reasoning behind governance decisions; justifications framed in terms reasonable persons across different comprehensive doctrines could accept.
\item[\textit{Adequate}] Democratic authorization present, with minor public-reasoning gaps that do not undermine the dimension's core function.
\item[\textit{Strained}] Significant gaps in public reasoning, including proprietary decision criteria, contested democratic basis, inaccessible justifications, or delegation to non-democratic actors.
\item[\textit{Failing}] No democratic authorization over an arrangement exercising quasi-governmental authority; decisions made by private actors without accountability to affected populations.
\end{description}

\paragraph{Accountability.}
\begin{description}
\item[\textit{Strong}] All three of~\citet{bovens2007analysing}'s conditions (information, debate, judgment) operational, with an independent forum and demonstrated enforcement.
\item[\textit{Adequate}] All three conditions present, with limited scope or capacity gaps that do not remove core function.
\item[\textit{Strained}] One or more conditions substantially impaired: information depends on self-reporting, enforcement authority constrained, or no structured debate forum.
\item[\textit{Failing}] One or more conditions absent: no external information flow, no competent forum, or no consequences.
\end{description}

\paragraph{Corrigibility.}
\begin{description}
\item[\textit{Strong}] Statutory external authority to correct, modify, or halt the system, demonstrably used.
\item[\textit{Adequate}] External correction authority exists with demonstrated use, though scope or enforcement are limited.
\item[\textit{Strained}] Correction happens primarily through voluntary compliance, internal mechanisms, or indirect channels. External authority is limited or contested.
\item[\textit{Failing}] No external mechanism exists to correct or halt the system.
\end{description}

\paragraph{Non-domination.}
\begin{description}
\item[\textit{Strong}] Clear contestatory rights with accessible procedures and demonstrated use.
\item[\textit{Adequate}] Contestatory rights present with minor practical limits on access or effectiveness.
\item[\textit{Strained}] Contestatory channels exist formally but face significant barriers (information asymmetry, resource asymmetry, procedural obstacles).
\item[\textit{Failing}] No effective contestatory channel exists for affected parties.
\end{description}

\paragraph{Subsidiarity.}
\begin{description}
\item[\textit{Strong}] Authority well-matched to scale, with functional implementation at each level and meaningful local adaptation.
\item[\textit{Adequate}] Authority appropriately allocated, with some implementation gaps at particular levels.
\item[\textit{Strained}] Authority allocation creates significant gaps: higher levels underpowered for the problem's scale, or lower levels unable to exercise assigned authority.
\item[\textit{Failing}] Authority fundamentally misallocated with clear harmful consequences; or, for fully private arrangements, no public authority over activity with substantial public impact.
\end{description}

\paragraph{Institutional resilience.}
\begin{description}
\item[\textit{Strong}] Multiple redundant mechanisms, institutional diversity, and demonstrated capacity under stress.
\item[\textit{Adequate}] Structural redundancy present; implementation capacity reasonable though not fully tested.
\item[\textit{Strained}] Structural redundancy exists but implementation capacity is limited or untested in ways that threaten function.
\item[\textit{Failing}] Single points of failure, no redundancy, demonstrated fragility.
\end{description}

\subsection*{Worked example: FDA accountability}

The FDA AI/ML case received a rating of \textit{strong} on accountability. The reasoning that produced this rating, traced against the indicators above, illustrates how each rating was determined.

\paragraph{Information.} FDA pre-market review produces public decision summaries, and the MAUDE adverse event database provides a post-market information channel external to the manufacturer. Both show that information about device conduct reaches the forum independently of the regulated entity. The information condition is operational.

\paragraph{Debate.} FDA advisory committees, congressional oversight hearings, and mandatory manufacturer reporting obligations create structured forums in which the FDA can be required to justify decisions and manufacturers can be required to explain device conduct. The debate condition is operational.

\paragraph{Judgment.} The FDA can issue recalls, require labeling changes, withdraw marketing authorization, and impose civil penalties. These are consequences imposable by a competent external body. The judgment condition is operational.

\paragraph{Aggregation.} All three Bovens conditions are operational, with an independent forum (the FDA) and demonstrated enforcement (recalls and post-market actions on authorized AI/ML devices). This matches the \textit{strong} indicator.

\paragraph{Counterpoint considered.} Post-market monitoring is structurally weaker than pre-market review, and accountability at the post-market stage shifts substantially to manufacturer self-reporting. A second analyst could weigh this asymmetry heavily enough to rate the dimension \textit{adequate} rather than \textit{strong}, on the grounds that the information condition is partially compromised after authorization. The rating reported in the main text is \textit{strong} because all three conditions function at the pre-market stage, which is where most regulatory contact occurs for current devices. The case discussion notes the post-market qualification explicitly.

The same reasoning process was applied to the other 35 dimension-case pairs, with the evidence cited inline in Section~\ref{sec:cases}.

\subsection*{Borderline ratings}

Three ratings are borderline between adjacent levels in ways that a second analyst could reasonably reverse.

\paragraph{EU AI Act, subsidiarity (adequate/strained).} The authority allocation between EU and member-state levels is defensible in principle, and the rating of \textit{adequate} reflects the view that the implementation gap (14 of 27 member states lacking designated authorities ahead of the deadline) is a capacity failure rather than an allocation failure. A second analyst could reasonably treat the implementation gap as substantive enough to warrant \textit{strained}, on the grounds that allocation without capacity is allocation in name only.

\paragraph{EU AI Act, institutional resilience (adequate/strained).} The multi-level architecture provides structural redundancy on paper, and the rating of \textit{adequate} reflects this structural reading. A more skeptical analyst could weigh the untested status of most bodies more heavily and rate the dimension \textit{strained}, on the grounds that redundancy is a property of institutions under stress, not of institutions before stress has arrived.

\paragraph{Voluntary commitments, subsidiarity (strained under attenuated reading).} The subsidiarity dimension presupposes that authority sits within a public governance hierarchy. Fully private arrangements do not fit this presupposition cleanly, and the rating of \textit{strained} rests on an attenuated reading in which the absence of any public authority over activity with substantial public impact counts as a subsidiarity failure by locating authority entirely outside public governance. A different analyst might treat the dimension as inapplicable to fully private arrangements and report N/A. The choice to rate rather than exempt preserves the six-by-six matrix and lets the paper make a specific point about what private governance does to subsidiarity; the alternative is noted here for transparency.

\bibliography{biblio}

\end{document}